\documentclass[12pt, a4paper]{article}
\usepackage{graphicx}
\usepackage[left=25mm, right=25mm, top=25mm, bottom=25mm]{geometry}

\usepackage[flushleft]{threeparttable} % http://ctan.org/pkg/threeparttable
\usepackage{paralist}
\usepackage{lipsum}
\usepackage{rotating}
\usepackage{mathrsfs}
\usepackage{scrextend}
\usepackage{tikz}
\usepackage{array}
\usepackage{enumerate}
\usepackage{ amsfonts,amsthm}
\usepackage{ multirow,appendix}
\usepackage{amsbsy}
\usepackage{authblk}
\usepackage{caption}
\usepackage{textcomp}
\usepackage{color}
\providecommand{\keywords}[1]{{\textit{Keywords and phrases.}} #1}
\usepackage{amsmath,float,fancyhdr,amsthm,amssymb,bm}

\renewcommand{\/}[1]{\bm{#1}}
\theoremstyle{plain}
\theoremstyle{definition}
\usepackage{natbib}
\newcommand{\nil}[1]{}
\newcommand{\atan}{\text{atan}}
\newcommand{\init}{\text{init}}
\newcommand{\var}{\text{var}}
\usepackage{fancyhdr}

\date{}
\begin{document}

\title{ Helix modelling through the Mardia-Holmes model framework and an extension of the Mardia-Holmes model }

\author[1]{Mai F Alfahad} 
\author[1]{John T Kent}
%\affiliation{Department of Statistics, University of Leeds, Leeds LS2 9JT, UK}
\author[1,2]{Kanti V Mardia}
%\affiliation{Department of Statistics, University of Leeds, Leeds LS2 9JT, UK}
%\affiliation{Department of Statistics, University of Oxford, UK}
\affil[1]{Department of Statistics, University of Leeds, Leeds LS2 9JT, UK}
\affil[2]{Department of Statistics, University of Oxford OX1 3LB, UK}
\affil[ ] {mmmfa@leeds.ac.uk, J.T.Kent@leeds.ac.uk, K.V.Mardia@leeds.ac.uk}
\maketitle

\begin{abstract}
  For noisy two-dimensional data, which are approximately uniformly
  distributed near the circumference of an ellipse,
  \citet{mardia1980holmes} developed a model to fit the  ellipse. In
  this paper we adapt their methodology to the analysis of helix data
  in three dimensions.  If the helix axis is known, then the
  Mardia-Holmes model for the circular case can be fitted after projecting the helix
  data onto the plane normal to the helix axis.  If the axis is
  unknown, an iterative algorithm has been developed to estimate the
  axis.  The methodology is illustrated using simulated protein
  $\alpha$-helices. We also give a multivariate version of the Mardia-Holmes model
which will be applicable for fitting an ellipsoid and in particular a cylinder. 
\end{abstract}

\maketitle %% required
\keywords{Fitted ellipse, Fitted circle, Principal component analysis, Helix axis,
Maximum likelihood, Least squares}
%-----------------------------------------------------
%                          Introduction
%-----------------------------------------------------
\section{Introduction}

A \emph{mathematical helix} is a curve in three dimensional space, of the form
\begin{equation}
\label{eq:math-helix}
\/f(t) = \begin{bmatrix}
r \cos t \\ r \sin t \\ ct \end{bmatrix}
\end{equation}
\citep[e.g.,][p. 16]{HELIXmodel},
augmented by an arbitrary rotation and shift in $\mathbb{R}^3$, as the
``time'' $t$ ranges through the real line.  This helix is called
``right-handed'' since when looked at from above, $x_1$ and $x_2$ move
in in a counter-clockwise direction around a circle as $t$ increases,
i.e. as the axis position $x_3$ gets closer to to observer.

A \emph{statistical helix} is obtained from (\ref{eq:math-helix}) by adding
noise at equally spaced time points $t_i = i\beta$ to give data
\begin{equation}
\label{eq:stat-helix1}
\/x_i = \/f(t_i) + \/\epsilon_i, \quad i=1, \ldots, n,
\end{equation}
where $\/\epsilon_i$ are small noise terms, typically modelled by
independent isotropic normal distributions,
$$
\/\epsilon_i \sim N_3(\/0, \sigma^2 I_3).
$$
The structural parameters of the helix data are
\begin{itemize}
\item the radius $r>0$;
\item the pitch $2\pi c$ (the amount of vertical movement after one rotation
around the helix); and
\item the turn angle $\beta$.
\end{itemize}

An important application of helix models is to secondary protein
structure, where a common structure is the right-handed
$\alpha$-helix, \citep[see e.g.][]{bio2014}.  A protein $\alpha$-helix
can treated as a data set of ``landmarks'' lying near a helix by
focusing on specific atoms such as $C_\alpha$ atoms.

An important task when presented with helix data is to estimate the
axis.  Various statistical methods have been proposed in the
literature. \cite{MardiaSSriram} used maximum likelihood estimates under various assumptions about the parameters. In particular if $\beta$ is known, it is also possible to use a modified least squares algorithm to compute the MLE; this particular algorithm was called OptLS by~\cite{Maiy} and this name will be used in this paper. There are many compositional methods to calculate the  axis; see for examples, \cite{aaqvist} and Rotfit by~\cite{Literature1}.

 In this paper, we develop a new method, by
adapting the \cite{mardia1980holmes} (M-H) model for data in the plane. The paper is laid out as follows.  In Section \ref{sec:MH}, the 
MH model for data in the plane is reviewed.  Then in Section \ref{sec:MH-axis}
the MH model is adapted to estimate the helix axis for three-dimensional data using a projection into the plane.
Section \ref{sec:sim} illustrates the use of the model on some 
simulated data.

\section{Mardia-Holmes model}
\label{sec:MH}
\cite{mardia1980holmes} (M-H) model was originally designed to analyze
megalithic data, in particular stones clustered uniformly around an
ellipse, or as a special case, a circle. The M-H model has several
parameters:
\begin{itemize}
\item a concentration parameter $\kappa >0$ describing how closely the
  data points are concentrated around an ellipse;
\item a location parameter in the plane $\/a=(a_1, a_2)^T$
  representing the centre of the ellipse; and
\item a $2 \times 2$ matrix $\Sigma$ used to specify the ellipse as a
  quadratic form,
\begin{equation}
\label{eq:ellipse}
(\/y-\/a)^T \Sigma^{-1} (\/y-\/a) = 1.
\end{equation}
\end{itemize}

The M-H model treats  $n$ data points in the plane $\/x_1, \ldots, \/x_n$
as independent observations from the density
\begin{equation}
\label{eqn:MH}
 f(\/y) = C(\kappa) |\Sigma|^{-1/2} 
\exp\{-\frac{1}{2} \kappa [(\/y-\/a)^{T} \Sigma^{-1} (\/y-\/a)-1]^2\},
\end{equation}
where $C(\kappa)=(\kappa / 2 \pi)^{1/2}/\{\pi \Phi(\kappa^{1/2})\}$ is
the normalization constant. This model has its mode on the
circumference of the ellipse, that is, the values of $\/y$ satisfying
(\ref{eq:ellipse}).

If $\Sigma=\rho^2 I_2,\  \rho^2>0$, and $I_2$ is the $2 \times 2$
identity matrix, then ellipse in (\ref{eq:ellipse}) reduces to a circle of
radius $\rho$. 

The circular version of the M-H model is the appropriate version for
helix data.  If data $\{\/x_i\}$ follow the helix model
(\ref{eq:math-helix})-(\ref{eq:stat-helix1}), then the projected
data
$$
\/y_i = \begin{bmatrix} y_{i1} \\ y_{i2} \end{bmatrix} =
\begin{bmatrix} x_{i1} \\ x_{i2} \end{bmatrix}
$$
will approximately follow the M-H model.  

The adjective ``approximately'' is needed for two reasons.  (i) The
angular parts of the helix data (i.e.
$\theta_i = \atan2(x_{i2},x_{i1})$, $i=1, \ldots, n$) are not i.i.d.;
the distribution of $\theta_i$ depends on $t_i$.  However, provided
$\beta/2\pi$ is not a simple fraction, we expect the $\theta_i$ to be
well spread around the circle.

(ii) The radial part of the  helix distribution (i.e. the
distribution of $(x_{i1}^2+x_{i2}^2)^{1/2}$) from
(\ref{eq:stat-helix1}) is not quite the same as the radial part of the
M-H distribution (i.e. the distribution of $(y_{i1}^2+y_{i2}^2)^{1/2}$
from (\ref{eqn:MH}).  However, the two radial distributions will be
very similar under high concentration, i.e. if $\kappa$ is large, by
matching the parameters $\kappa = 1/\sigma^2$.

Estimation in the M-H model can be done by maximum likelihood. However,
since the MLEs do not exist in closed form, an iterative algorithm is needed.
The simplest procedure is to choose plausible initial estimates and then
to use a black box optimization algorithm (e.g. the function \texttt{nlm} in 
R) to carry out the maximization.

Here is a set of choices of initial estimates for the circular case,
given data $\{\/y_i\}$:
\begin{itemize}
\item Estimate $\/a$ by $\hat{\/a}_\init$, the vector mean of the data.
\item Estimate $\rho$ by $\hat{\rho}_\init$, the average distance
  between the data and $\hat{\/a}_\init$, i.e.
  $n^{-1} \sum |\/y_i - \hat{\/a}_\init|$.

\item Estimate $\kappa$ by the reciprocal of the sample variance of the radial
part of the centered data,
$$
1/\hat{\kappa}_\init = \var\{|\/y_i - \hat{\/a}_\init|\}.
$$
\end{itemize}
In the \texttt{nlm} procedure it is convenient to use unconstrained parameters.
Hence we  work with $\eta=\log(\kappa)$ and $\tau=\log(\rho)$.
The output of this procedure is a set of estimates and a value MLL, say, for
the maximized log likelihood.

\section{Estimating the helix axis using the M-H model}
\label{sec:MH-axis}
If an arbitrary rotation and shift are added to the mathematical helix model
(\ref{eq:math-helix}), then it is convenient to write the model in the form
\begin{equation}
\label{eq:stat-helix2}
\/x_i = r (\cos t_i) \/u + r (\sin t_i) \/v + c t_i \/w + \/b + \/\varepsilon_i ,
 \end{equation}
 where $\/u, \/v, \/w$ are three-dimensional orthonormal vectors.  In
 particular,  $\/w$ is the helix axis. Further,
 $R = [\/u \ \/v \ \/w]$ is a $3 \times 3$ rotation matrix.  The
 vector $\/b$ represents the shift term.  The rotated errors
 $\/\varepsilon_i = R \/\epsilon_i$ still follow independent isotropic normal
 distributions, $\/\varepsilon_i \sim N_3(\/0, \sigma^2 I_3)$.

If $R$ is a possible estimate of the rotation matrix, then 
the first two components of the rotated data
$$
\/y_i = \begin{bmatrix} y_{i1} \\ y_{i2} \end{bmatrix} =
\begin{bmatrix} (R^T \/x_i)_1 \\ (R^T \/x_i)_2 \end{bmatrix}
$$
will approximately follow the M-H model.  Since the fit of the M-H model is
invariant under rotations in the plane, the maximized M-H log likelihood
for the $\{\/y_i\}$ depends only on the third column of $R$, i.e. $\/w$,
and not on the relative orientation of the first two columns.  Hence write
$MLL(\/w)$ for the maximized M-H log likelihood, depending on the choice
of helix axis $\/w$.

To estimate $\/w$ we maximize $MLL(\/w)$ over $\/w$.  As in the
last section it is necessary to use an interative numerical method
such as the R routine \texttt{nlm}, starting from an initial
estimate of $\/w$.

A suitable initial estimate $\/w_\text{init}$ can be found using e.g.
modified principal component analysis or OptLS \citep{Maiy} or Rotfit
\citep{Literature1}.

A unit vector is a constrained vector in three dimensions.  For
optimization purposes, it is helpful to represent it using
unconstrained two-dimensional coordinates.  For example, once an
initial estimate has been selected, we can rotate the data so that the
initial estimate points to the north pole, $[0 \ 0 \ 1]$ and represent
deviations about the north pole using stereographic coordinates
$\/p=(p_1,p_2)^T$, say, where $\/w$ can be written in terms of $\/p$
as follows:
\begin{align*}
w_1&= \frac{2{p}_{1}}{1+{p}_{1}^{2}+{p}_{2}^{2}},\\
w_2&=\frac{{2p}_{2}}{1+{p}_{1}^{2}+{p}_{2}^{2}},\\
w_3&=\frac{-1+{p}_{1}^{2}+{p}_{2}^{2}}{1+{p}_{1}^{2}+{p}_{2}^{2}}.
\end{align*}

Then define a function
\begin{equation}
\label{eq:outer}
f(p_1,p_2)= MLL(\/w)
\end{equation}
and maximize this function numerically starting at $\/p=\/0$.

Note that the M-H procedure involves a nested use of numerical optimization.
At the inner level, numerical optimization is used to maximize the M-H
log likelihood, assuming the helix axis $\/w$ is given, yielding a maximized
log likelihood $MLL(\/w)$.  At the outer level,
we maximize (\ref{eq:outer}) over $\/p$, i.e. over the choice of $\/w$.

We now apply the method to estimate the axis for two real helices (Helices 7 and 8  from Mardia et al (2018)).  For Helix 7 the estimated  axis  is $(0.591, -0.795, 0.133)^T$  and  for Helix 8 the estimated  axis is $(0.336, 0.516, -0.788)^T$. Their estimates from OptLS  are respectively $(0.601, -0.789, 0.129)^T $ and $(0.318, 0.537, -0.780)^T$.  The cosine and their angle  for the two cases are
0.9999315, $\theta$ = 0.01170463; 0.9995821, $\theta$ = 0.0289111
These indicate that for the two cases, these estimates are very  similar.  The next section, examines their mean square error  through a simulation study.

\section{Simulation}
\label{sec:sim}
In this section, we illustrate our M-H procedure for estimating the
helix axis on 100 simulated helices that mimic a protein
$\alpha$-helix for different choices of sample size $n$ and parameter
values $r$, $c$, and $\sigma^2$ (with $\beta = 2\pi/3.6$)and compare with OptLS procedure. For
more details of protein $\alpha$-helix see \cite{mardia2013protein},
\cite{protein} and \cite{protein2}. We have 100 estimates of the helix
axis $\hat{\/w}_{\text{M-H}, i}$ by the M-H procedure and
$\hat{\/w}_{\text{Opt}, i}$ by the OptLS method. To calculate the
accuracy of these estimates, we define the mean square errors (MSEs)
in terms of the means of the inner products,
$$
\text{MSE}_{\text{M-H}} = 1-\hat{\bar{\/w}}_{\text{M-H}}^{T} \/w_0, \quad
\text{MSE}_{\text{Opt}} = 1-\hat{\bar{\/w}}_{\text{Opt}}^{T} \/w_0,
$$
where the inner products are sample means,
$$\
\hat{\bar{\/w}}_{\text{M-H}}=\frac{1}{100}\sum_{i=1}^{100}
\hat{\/w}_{\text{M-H}, i}, \quad 
\hat{\bar{\/w}}_{\text{Opt}}=\frac{1}{100}\sum_{i=1}^{100}
\hat{\/w}_{\text{Opt}, i},
$$
and $\/w_0=(0, 0, 1)^T$ is the
axis pointing to north pole. Let $\theta$ be the angle between the
estimated axis $\hat{\bar{\/w}}_{\text{M-H}}$ (or
$\hat{\bar{\/w}}_{\text{Opt}}$) and $\/w_0$. If this angle vanishes, $\theta=0$,
then the estimated axis is a perfect fit, then the inner product is 1 
\citep{kinkangle}, so that  the MSE is equal to zero.

We illustrate the algorithm with an example of one simulated dataset that mimics a long protein $\alpha$ helix, where $n=30, r=2.3, c=5.4/(2\pi), \beta=2\pi/3.6, \sigma^2 = 0.001$, and the true axis is $\/w_0=(0, 0, 1)^T$.  We also estimate the axis of this dataset by OptLS. The estimated helix axis by M-H procedure is $\hat{\/w}_{M-H}=(-5.9 \times 10^{-4}, 3.2 \times 10^{-4}, 0.9999999)^T$ and the estimated helix axis by OptLS is  $\hat{\/w}_{Opt}=(-3.8 \times 10^{-4}, 3.0 \times 10^{-4}, 0.9999998)^T$. The MSE by M-H procedure is $2.3 \times 10^{-7}$ and the MSE by OptLS is $1.1 \times 10^{-7}$. This result shows that OptLS is more accurate than M-H algorithm as the MSE is smaller by a factor of two.

Table \ref{table:meanVar2} shows the MSE for six different simulated
datasets.  
These data sets have been constructed with a variety of
parameter choices.  Set 2 mimics a long protein $\alpha$ helix
($n=30$) and set 3 mimics short protein $\alpha$ helix ($n=12$). The
remaining data sets are modified versions of these sets.  In particular,
the error variance has been decreased for set 1 from 
$\sigma^2=0.05$ to $\sigma^2 = 0.001$ and increased for set 4 from $\sigma^2=0.05$ to $\sigma^2 = 0.10$. Sets 5 and 6 are  fatter
helices (changing $r=2.3$ to $r=7$); in addition for set 5, the pitch
parameter $c$ has been reduced (from $c=5.4/(2\pi)= 0.859$ to $c=0.63/(2\pi)=.1$). The parameter $\beta=2\pi/3.6$ is fixed for all these sets.

In each case the MSE of M-H is at least a factor of two larger than the MSE of OptLS. From
this result we conclude that the OptLS is generally much more accurate
than the M-H procedure. This is partly expected for the reasons (i) and ( ii) given in Section 2. Furhter , we are comparing only these two methods but of course there are other methods, mainly computational.  

\begin{table}[H]
\centering
\caption{Comparison between M-H and OptLS procedures by the mean square error.}
\label{table:meanVar2}
\begin{tabular}{lllllll}
\hline\hline
\rule{0pt}{2.5ex}  &\rule{0pt}{3ex} set1 & set 2& set 3& set 4& set 5 & set 6\\
\hline\hline
\rule{0pt}{2.5ex} n    &   \rule{0pt}{3ex} 30& 30& 12 & 12 & 12& 12\\
\rule{0pt}{2.5ex} r     &   \rule{0pt}{3ex} 2.3& 2.3&2.3 & 2.3 & 7 & 7 \\
\rule{0pt}{2.5ex} c     &   \rule{0pt}{3ex}  $\frac{5.4}{(2\pi)}$&$\frac{5.4}{(2\pi)}$&$\frac{5.4}{(2\pi)}$&$\frac{5.4}{(2\pi)}$&$\frac{0.63}{(2\pi)}$&$\frac{5.4}{(2\pi)}$\\
\rule{0pt}{3ex}$\sigma^2$  &   \rule{0pt}{3ex}  0.001&0.05&0.05&0.1 &0.05&0.05\\
\rule{0pt}{2.5ex} M-H     &   \rule{0pt}{3ex}  $2.8\times 10^{-7}$&$1.5\times 10^{-5}$&$2.4\times 10^{-4}$&$4.5\times 10^{-4}$&$1.2\times 10^{-2}$&$2.3\times 10^{-4}$\\
\rule{0pt}{2.5ex} OptLS        &\rule{0pt}{3ex}  $1.2\times 10^{-7}$&$0.5\times 10^{-5}$&$1.4\times 10^{-4}$& $2.8\times 10^{-4}$ &$0.01\times 10^{-2}$&$0.8\times 10^{-4}$\\ \hline
\end{tabular}
\end{table}

\section {Appendix: Extension of Mardia and Holmes Model}
 Mardia and Holmes's bivariate model  can be extended to any dimension on replacing the ellipse by an ellipsoid.  Namely, let X be a random vector in   $d$ dimension then their  distribution   has the probability density function (p.d.f.)
\begin{equation}
 f(x; \mu, \Sigma,\kappa) = C(\kappa) |\Sigma|^{-1/2}   exp \{-\frac{1}{2}
 \kappa ((x-{\mu}) ^T {\Sigma}^{-1}(x-{\mu})  -1)^2\} 
\end{equation}
where $x$ is  dimension $d$  the concentration parameter$\kappa>0$; $\mu$ takes any value in $R^d$ and $\Sigma$ is a positive definite matrix.
It turns out that the normalizing constant is given by a parabolic cylindrical  function and is given  below. For the right cylinder, the two small eigenvalues are equal a priori. The axis is the z-axis if we take the largest eigenvalue as the third one; 
that is a thin and long ellipsoid so may be relevant "tangentially" though the multivariate Mardia-Holmes  model will be useful for any inference related to fitting an ellipsoid.

We will use the  general summary of  elliptic family  by Azzilini(2014,pp.168-169) of which this is a member but not studied.
Let us write 
$$ r^2 =(x-{\mu}) ^T {\Sigma}^{-1}(x-{\mu}). $$ 
Then the pdf can be rewritten as 
\begin{equation}
 f(x; \mu, \Sigma,\kappa) = C(\kappa) |\Sigma|^{-1/2} p(r^2)  
\end{equation}
where 
$$C(\kappa) =\Gamma(d/2)/(2\pi^{d/2}b(\kappa)),$$ 
with 
\begin{center}
$b(\kappa)= \int_0^\infty r^{d-1} p(r^2)\ dr$ and
$p(r^2)=\exp\{-\frac{1}{2}
 \kappa (r^2  -1)^2\}$.
\end{center} 
 
We need now to evaluate $b(\kappa)$ given by 
 
$$b(\kappa)= \int_0^\infty r^{d-1} \{\exp- \frac{1}{2}
 \kappa (r^2  -1)^2\}\ dr.$$
 In fact, it can be expressed in terms of the parabolic cylindrical function defined by (see, Abramowitz and Stegun, 1965, Chapter 9, p.688).
 $$U(a,z) = \frac{1}{\Gamma(a + \frac{1}{2} )}\exp \left(-\frac{1}{4}z^2 \right) \int_0^{\infty} s^{a-\frac{1}{2}} \exp \left(-\frac{1}{2}s^2 - zs \right) ds .$$
 We have 
\begin{equation}
 b(\kappa)=  \frac{\Gamma(a + \frac{1}{2}) \exp(-\frac{\kappa}{4})} {2 \kappa^\frac{2a+1}{4}}  U(a,-\sqrt\kappa),
\end{equation}
where $a=(d-1)/2$. Note that if $d=2n$ then $a=n-\frac{1}{2}$ and if $d=2n+1$ then $a=n$. For $d=3$, we have $a=1$.

We now summaries a few properties
The mode of the distribution is given by
$$(x-{\mu}) ^T {\Sigma}^{-1}(x-{\mu})  =1.$$  
 Further,
\begin{equation}
 E(X)=\mu ,\;\; E (X-\mu)(X-\mu)^T  = \alpha\Sigma,
\end{equation}
where $\alpha$ is a function of $\kappa$.
The most general equation of ellipsoid is given by
$$(x-{\mu}) ^T {\Sigma}^{-1}(x-{\mu})  =1 $$  
so 1 in the LHS does NOT  need any adjustments. It can be noted that the model 
works to fit a general ellipsoid  when there is a very high probability of  concentration around 
$$(x-{\mu}) ^T {\Sigma}^{-1}(x-{\mu})  =1$$ where ${\Sigma}$  is positive definite. \nocite{abramowitzhandbook}\nocite{azzalini2014skew}

%
%\renewcommand{\&}{and}
%\bibliographystyle{apalike2}
%\bibliography{MHref}
\end{document}